%% file: conference_101719.tex
\documentclass[conference]{IEEEtran}
\IEEEoverridecommandlockouts
\usepackage{cite}
\usepackage{amsmath,amssymb,amsfonts}
\usepackage{algorithmic}
\usepackage[acronym]{glossaries}
\usepackage{graphicx}
\usepackage{textcomp}
\usepackage{xcolor}

\loadglsentries{abbreviations}

\def\BibTeX{{\rm B\kern-.05em{\sc i\kern-.025em b}\kern-.08em
    T\kern-.1667em\lower.7ex\hbox{E}\kern-.125emX}}

\begin{document}

\title{UltraScatter: Ray-Based Simulation of Ultrasound Scattering
}
\makeatletter
\newcommand{\linebreakand}{%
  \end{@IEEEauthorhalign}
  \hfill\mbox{}\par
  \mbox{}\hfill\begin{@IEEEauthorhalign}
}
\makeatother

\author{
\IEEEauthorblockN{Felix Duelmer\IEEEauthorrefmark{1}\IEEEauthorrefmark{2}\IEEEauthorrefmark{3}\IEEEauthorrefmark{4}, Mohammad Farid Azampour\IEEEauthorrefmark{1}\IEEEauthorrefmark{2}, Nassir Navab\IEEEauthorrefmark{1}\IEEEauthorrefmark{2}}
\IEEEauthorblockA{\IEEEauthorrefmark{1}Chair for Computer Aided Medical Procedures (CAMP), School of
Computation, Information and Technology, \\ Technical University of Munich, Munich, Germany}
\IEEEauthorblockA{\IEEEauthorrefmark{2}Munich Center for Machine Learning (MCML), Munich, Germany}
\IEEEauthorblockA{\IEEEauthorrefmark{3}Institute of Biological and Medical Imaging, Bioengineering Center, Helmholtz Zentrum München, Neuherberg, Germany}
\IEEEauthorblockA{\IEEEauthorrefmark{4}Chair of Biological Imaging, Central Institute for Translational Cancer Research (TranslaTUM), \\ School of Medicine and Health \& School of Computation,
Information and Technology,\\ Technical University of Munich, Munich, Germany\\
Email: felix.duelmer@tum.de}
}

\maketitle

\begin{abstract}
Traditional ultrasound simulation methods solve wave equations numerically, achieving high accuracy but at substantial computational cost. Faster alternatives based on convolution with precomputed impulse responses remain relatively slow, often requiring several minutes to generate a full B-mode image. We introduce UltraScatter, a probabilistic ray tracing framework that models ultrasound scattering efficiently and realistically. Tissue is represented as a volumetric field of scattering probability and scattering amplitude, and ray interactions are simulated via free-flight delta tracking. Scattered rays are traced to the transducer, with phase information incorporated through a linear time-of-flight model. Integrated with plane-wave imaging and beamforming, our parallelized ray tracing architecture produces B-mode images within seconds. Validation with phantom data shows realistic speckle and inclusion patterns, positioning UltraScatter as a scalable alternative to wave-based methods.
\end{abstract}


\section{Introduction}

Ultrasound simulation is essential for enhancing reconstruction algorithms, optimizing transducer designs, and training machine-learning models under controlled conditions. Conventional wave-based solvers such as FDTD \cite{hallaj1999fdtd} and k-space pseudospectral methods \cite{treeby2010k} are accurate but computationally demanding and difficult to scale. Quicker methods like Field II \cite{jensen1991model} simulate RF signals by convolving precomputed spatial impulse responses with the excitation pulse at each scatterer location. SIMUS  \cite{garcia2022simus}, on the other hand, computes RF signals by applying delay-and-sum operations based on the geometric round-trip distances between scatterers and transducer elements. Despite being more efficient than full-wave solvers, these simulators still require several minutes to create a single B-mode image, underscoring the need for approaches that combine physical realism with rapid computation.
 
\Gls{crt} has recently emerged as a fast alternative to wave-based ultrasound simulation. Representative implementations include \cite{mattausch2018realistic, amadou2024cardiac, burger2012real, wein2007simulation, shams2008real, duelmer2025ultraray}. These techniques approximate acoustic wave propagation by tracing large ensembles of rays that reflect and refract at macroscopic boundaries. Fine-scale scattering is then modelled by convolving a separable \gls{psf} with a predefined distribution of point scatterers, following Gao \emph{et al.} \cite{gao2009fast}. Scattering probability and amplitude parameters govern the echo strength, but because they are typically drawn at random, the resulting scattering field is inherently stochastic.

Concurrently, the computer-graphics community has advanced realistic image synthesis through physically based rendering (PBR) \cite{pharr2023physically}. In PBR, scenes are defined by light sources, sensors, and objects with material properties. Light transport is then simulated by tracing rays and determining whether a path connects a sensor (camera) to an emitter via scene interactions. These interactions set the radiance that contributes to each pixel. Techniques such as ray marching \cite{kajiya1984ray}, photon mapping \cite{veach1998robust}, and delta tracking \cite{novak2014residual} address light propagation in heterogeneous media, and highly optimized implementations of these methods now power modern photorealistic renderers like Mitsuba 3 \cite{Mitsuba3}.

Exploiting the highly optimized, CUDA-accelerated ray-tracing algorithms developed in computer graphics, we embed a ray tracing algorithm within the conventional ultrasound image-formation pipeline. Our main contributions are (i) a modular, high-performance framework that models attenuation, absorption, and multiple scattering in participating media, (ii) an emitter sampling strategy that connects every scene interaction with all the transducer elements, and (iii) a full transmit–receive beamforming chain that converts the simulated echoes into B-mode images directly.

\section{Methodology}

Figure~\ref{fig:pipeline_overview} gives an overview of the proposed simulation pipeline to create  \gls{rf} data.  
Starting from a label map, we assign scattering properties to each tissue class. Using a Monte-Carlo ray tracing scheme, we model pressure-wave emission, scattering, and attenuation within the medium. Echoes returned to the transducer are written to element-specific RF buffers and then processed by a conventional digital beamformer to produce the final B-mode image.  
The next section details the ray-tracing stage, while Sec.~\ref{sec:implementation_details} outlines implementation specifics and beamforming parameters.

\begin{figure}[htbp]
    \centering
    \includegraphics[width=0.48\textwidth]{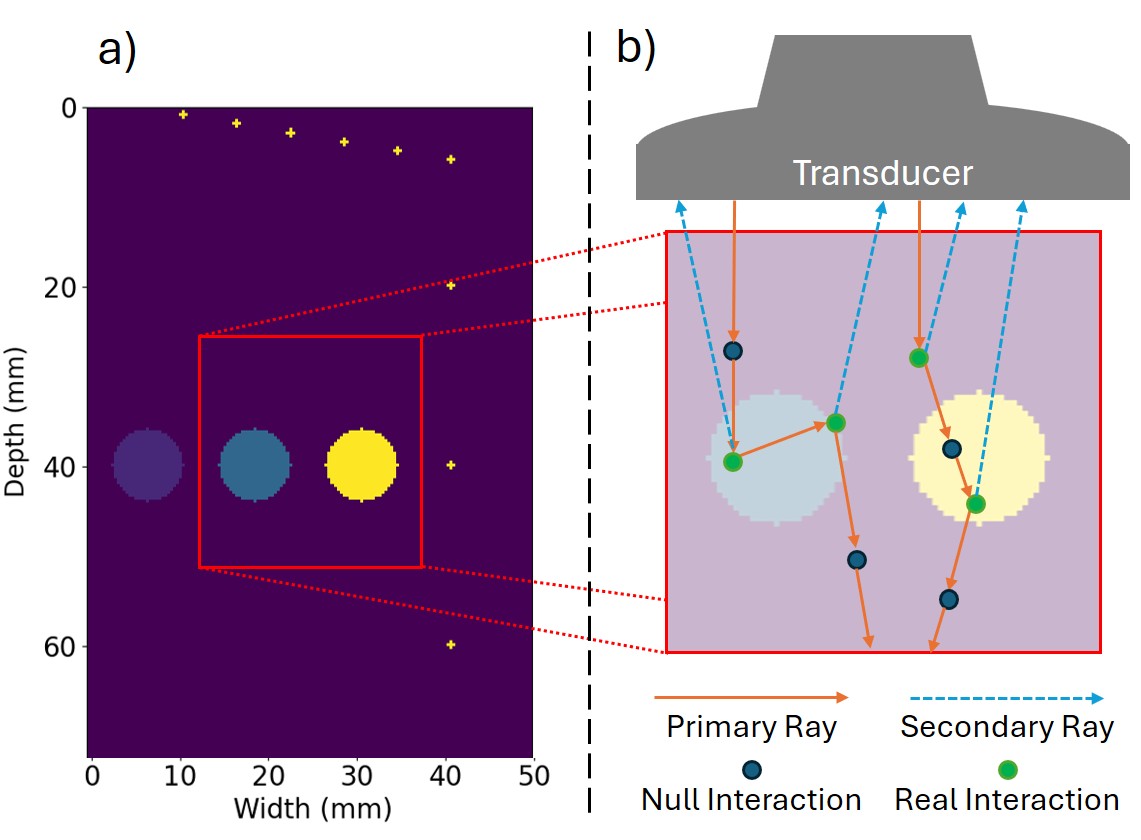}
    \caption{Overview of the ray-tracing pipeline: (a) label map indicating the different scattering regions, 
(b) example primary rays, experiencing  null interactions and real interactions. After each real interaction and a subsequent evaluation for scattering or absorption, secondary rays are launched toward the transducer. Only one secondary ray per scattering event is depicted for improved clarity.}
    \label{fig:pipeline_overview}
\end{figure}

\subsection{Ray Tracing}\label{sec:ray_tracing}

When simulating ultrasound, we need to track pressure waves from the ultrasound transducer through the scene (i.e., tissue) and observe the backscattered pressure waves. In theory, this whole pipeline can be represented as a multitude of nested integrals (over surface, direction, etc.) that all contribute to the final received pressure at a time $t$ following an excitation pulse. However, this quickly becomes infeasible. We, therefore, similarly to \cite{duelmer2025ultraray, amadou2024cardiac, mattausch2018realistic}, adapt a Monte-Carlo ray tracing scheme to replace the integrals with sampling strategies weighted by their respective probabilities. This effectively replaces a wave with a wavefront of rays, which can  be calculated way more effectively on modern-day hardware.

Following Duelmer \emph{et al.} \cite{duelmer2025ultraray} we describe the pressure signal $P$ that is arriving at the transducer element $e$ at time $t$ as:

\begin{equation}
    P(e, t) = \int_{\Omega} \int_{A} P_i(\mathbf{x}, t, \omega_i) \, f_d(\omega_i) \, d\omega \, da.
\end{equation}
This equation integrates the incoming pressure $P_i$ from position $\mathbf{x}$ in the scene, coming from direction $\omega_i$. By weighting this contribution with a directivity function $f_d$, we emulate a baffle in a real transducer.

At each interaction point $\mathbf{x}$ in the medium, an incoming wave with pressure $P_i(\mathbf{x}, t, \omega_i)$ arriving from direction $\omega_i$ is scattered into an outgoing direction $\omega_o$. The scattered pressure is modeled as:
\begin{equation}
P_{\text{scattered}}(\mathbf{x}, t, \omega_o) = \int_{\Omega} a(\mathbf{x}) \, p(\omega_i, \omega_o) \, P_i(\mathbf{x}, t, \omega_i) \, d\omega_i,
\end{equation}
where $a(\mathbf{x})$ denotes the scattering amplitude and $p(\omega_i, \omega_o)$ is the phase function that governs angular redistribution.

An interaction event occurs following the principles of free-flight delta tracking with null interactions, also known as Woodcock tracking \cite{ woodcock1965techniques,novak2014residual}. Scattering is determined by an extinction coefficient $\sigma_t$, which is proportional to the scattering probability, and the scattering amplitude, which defines the proportion of scattered versus absorbed energy.

To determine whether and where an interaction occurs, we first sample a free-flight distance $s$ that a ray travels before an interaction is triggered. This sampling is based on the extinction majorant $\mu$, which is the maximum extinction value across the medium. A random number $\xi \sim \mathcal{U}(0,1)$ is drawn, and the distance $s$ is computed as:
\begin{equation}
s = s_{\min} - \frac{1}{\mu} \ln(1 - \xi)
\end{equation}

Once a candidate interaction position $\mathbf{x}$ is sampled, we determine whether it constitutes a real or null interaction by comparing another random number $\xi'$ to the ratio of the local extinction coefficient and the majorant:
\begin{equation}
\xi' < \frac{\sigma_t(\mathbf{x})}{\mu} \quad \Rightarrow \quad \text{real interaction, else null interaction}
\end{equation}

In the case of a real interaction, the scattering amplitude at $\mathbf{x}$ governs whether the ray is absorbed or scattered. If scattering occurs, a new direction $\omega_o$ is sampled according to the phase function $p(\omega_i, \omega_o)$, and the scattered pressure is updated accordingly.

\begin{figure*}[htbp]
    \centering
    \includegraphics[width=\textwidth]{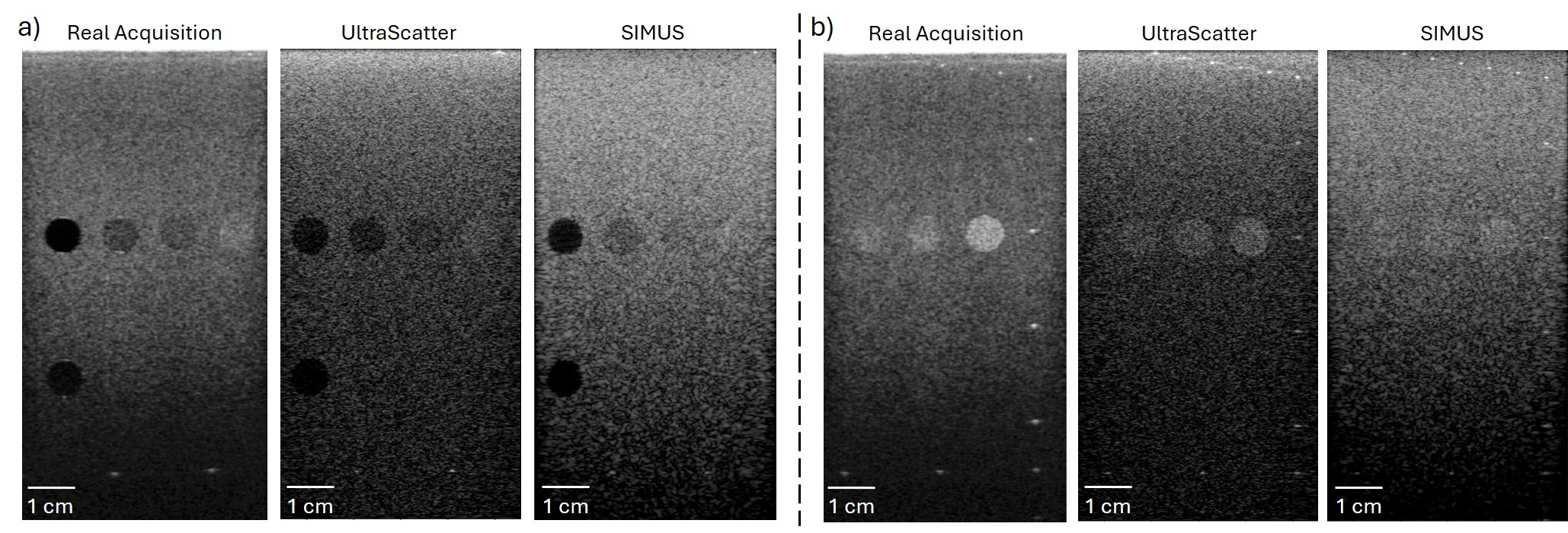}
    \caption{B-mode slices of the CIRS 054 GS phantom acquired (from left to right) with the clinical scanner, simulated with UltraScatter, and simulated with SIMUS. Two probe positions are shown: (a) first lateral view and (b) second lateral view. UltraScatter slightly blurs and distorts the inclusions closest to the transducer surface, but resolves the deeper cylinders more sharply than SIMUS. }
    \label{fig:results_comparison}
\end{figure*}

To increase the likelihood that rays ultimately reach the transducer, we employ an emitter-sampling strategy in which every transducer element is sampled at each scattering event, contrary to the selective approach of Duelmer \emph{et al.}~\cite{duelmer2025ultraray}. In our experiments, this comprehensive sampling consistently produced cleaner images with sharper structural boundaries. At each interaction point we therefore launch a set of secondary rays, one directed toward each transducer element. These secondary rays undergo the same physical processes (e.g.\ attenuation and phase delay) as the primary rays, but they are constrained to a single, predetermined direction.

When a secondary ray reaches its target element, its total path length is used to compute that element’s contribution to the \gls{rf} signal. Specifically, the ray’s remaining pressure and travelled distance define an axial pulse, a sinusoid modulated by an envelope, following \cite{duelmer2025ultraray}, which is stored in memory. The chosen sampling frequency then determines the discrete time instants at which this pulse is recorded.

Because our received signal is computed using simple phase summation based on distance, we rely on the Fraunhofer far-field approximation, which assumes planar wavefronts. To ensure this assumption holds, we subdivide each transducer element into $\nu$ identical sub-elements, following the approach in \cite{garcia2022simus}. This reduces the effective aperture size of each emitter, thereby shifting the near-to-far field transition closer to the transducer and enforcing the validity of the far-field condition throughout the imaging region. Specifically, we require:

\begin{equation}
    \nu = \left\lceil \frac{2b}{\lambda_{\min}} \right\rceil
\end{equation}
and split each element accordingly. The sub-elements are then treated as independent emitters/receivers in the subsequent phase-summation step. $b$ denotes the width of a single transducer element in the azimuthal direction, and $\lambda_{\min}$ is the smallest wavelength in the transmitted pulse.

We map a relative dB field \(\Delta_{\text{dB}}(\mathbf{x})\) to the scattering amplitude \(a\) and then invert the mapping (via the log-scaled relation \(I = I_{\max}[1+\tfrac{20}{\text{dB}_{\text{range}}}\log_{10}(a)]\)) to recover a spatially varying intensity \(I(\mathbf{x})\). Using the same \(a(\mathbf{x})\) both as the scattering amplitude and as the sampling probability automatically suppresses contributions from regions that contain few scatterers, yielding a realistically heterogeneous medium with balanced variance.

In practice we swap each multi-dimensional integral for a Monte-Carlo sum of weighted ray contributions:
\begin{equation}
\int_{\Omega}\!\int_{A} f(\mathbf{x},\omega)\,d\omega\,dA \;\approx\; \frac{1}{N}\sum_{k=1}^{N}\frac{f(\mathbf{x}_k,\omega_k)}{p(\mathbf{x}_k,\omega_k)} .
\end{equation}
Here \((\mathbf{x}_k,\omega_k)\) are samples drawn from the density \(p\), and every ray carries the weight \(1/p(\mathbf{x}_k,\omega_k)\). This mirrors the approach of Duelmer \emph{et~al.}~\cite{duelmer2025ultraray}, augmented with our full-element sampling, null-collision tracking, and far-field sub-element split.

\subsection{Implementation Details}\label{sec:implementation_details}

Our ray tracing implementation builds upon Mitsuba 3 \cite{Mitsuba3}, a physics-based rendering framework originally developed for simulating forward and inverse light transport in natural images. Mitsuba is written in C++ with Python bindings and provides a modular architecture that supports both CPU and GPU rendering backends. For this work, we utilize the GPU-based variant, which is accelerated via NVIDIA's OptiX framework \cite{parker2010optix}. To adapt Mitsuba for ultrasound simulation, we developed custom modules in both C++ and Python, extending core components such as the emitter, sensor, film, reconstruction filter, memory block, and integrator to model acoustic wave propagation and interaction.

Beamforming and signal processing are performed using the Ultraspy library \cite{ecarlat2023get}. All ray tracing simulations and comparison simulations were executed on a workstation equipped with an Intel Core i7-12700 CPU (20 threads) and an NVIDIA RTX 4070 Ti GPU.

\section{Results}

We compared UltraScatter with a clinical scan from a Siemens Acuson Juniper machine that used a 12L3 linear array containing 192 elements and operating between 2.9 and 11.5 MHz. The acquisition ran at a centre frequency of 6.2 MHz with an imaging depth of 10 cm and displayed a dynamic range of 60 dB. SIMUS, a frequency domain time harmonic simulator that relies on far field and paraxial assumptions, served as a second reference \cite{garcia2022simus}. Both SIMUS and UltraScatter were set up to reproduce the physical probe and both used a single plane wave transmit at normal incidence, a sampling rate of 25 MHz, and receive beamforming with an F-number of 1.0, together with a Hanning window. We emitted one hundred thousand rays for every sub-element.

The evaluation used the CIRS general-purpose phantom (Model 054GS). We reconstructed its interior from the manufacturer’s specifications and adjusted missing parameters until the simulated reference matched the clinical B-mode. As outlined in Sec. \ref{sec:ray_tracing}, the label volume was converted to scattering-probability and amplitude fields for both SIMUS and UltraScatter. To enrich speckle, each voxel was multiplied by a uniform random value in [0, 1]. For SIMUS, we then randomly culled voxels until roughly $1 × 10^5$ scatterers remained, a density that produced a realistic texture in qualitative tests.

When the two phantom views illustrated in Fig.\,\ref{fig:results_comparison} were rendered, SIMUS, even when run with multiprocessing, required \(634 \pm 3\)\,s of wall-clock time on our test workstation, whereas UltraScatter completed the same task in \(9.3 \pm 0.8\)\,s, representing a speed-up of nearly a factor of seventy.  Visually, the three B-mode images share the overall anatomy: the background speckle statistics are similar, and the distal shadowing mirrors the real scan.  Nevertheless, subtle differences are evident. UltraScatter tends to over-illuminate regions immediately beneath the probe, a consequence of residual overestimation of the near-field gain, while SIMUS reproduces the axial intensity fall-off of the clinical system more faithfully but renders inclusion edges with slightly less definition.  UltraScatter, on the other hand, produces sharper inclusion boundaries, which we attribute to its full-element emitter sampling strategy. Taken together, the experiment shows that UltraScatter attains an image quality comparable to SIMUS and the real scanner, yet accomplishes this with a runtime two orders of magnitude shorter, making real-time or even interactive simulation scenarios feasible on commodity hardware.

\section{Discussion and Conclusion}

UltraScatter is a ray-tracing ultrasound simulator that replaces frequency-domain solvers with Monte-Carlo path sampling. It models scattering in heterogeneous tissue, traces each ray through a three-dimensional volume, and feeds the resulting pressures into a conventional delay-and-sum beamformer that produces synthetic B-mode frames. On identical hardware, the program renders a typical CIRS phantom view in about nine seconds, whereas SIMUS needs more than ten minutes. The speed-up results from the parallel nature of ray evaluation and from the absence of large Fourier transforms. Runtime can be shortened even further by lowering the ray count, reducing the temporal sampling rate, or choosing a lower centre frequency. Each of these changes has a predictable cost in image fidelity because fewer paths are averaged and high-frequency content is suppressed.

Speckle statistics are driven by random sampling. Two independent runs, therefore, yield different patterns, and the texture decorrelates as soon as the virtual probe translates laterally. Future work is necessary to introduce correlated random sequences or shared seed maps so that speckle remains stable over small probe motions, preserving temporal coherence in dynamic scenes.

The current prototype launches rays from a single axial line into a cubic grid. A more realistic configuration would transmit and receive across the full elevational aperture of the transducer. Implementing this change will require elevational focusing, achieved either by dynamic receive delays or by casting additional rays. The added cost is expected to sharpen the point-spread function in three dimensions.
Ray tracing also makes it straightforward to assign each voxel its own density, sound speed, and attenuation. UltraScatter, therefore, has the potential of modeling macroscopic refraction, phase aberration, and even nonlinear phenomena such as harmonic generation, none of which fit within the weak-scatterer assumption used by many linear simulators.

In its present form, UltraScatter already produces B-mode images that closely resemble clinical scans while running roughly seventy times faster than a traditional frequency-domain code. The architecture is modular and leaves room for forthcoming improvements in speckle consistency, elevational resolution, and support for dynamic probe motion.

\bibliographystyle{IEEEtran}
\bibliography{references}

\end{document}

%% file: abbreviations.tex
\newacronym{us}{US}{Ultrasound}
\newacronym{das}{DAS}{Delay-and-Sum}
\newacronym{rf}{RF}{radio-frequency}
\newacronym{psf}{PSF}{point spread function}
\newacronym{crt}{CRT}{convolutional ray tracing}
\newacronym{pbr}{PBR}{physics-based rendering}